\documentclass[letterpaper,twocolumn,10pt]{article}
\usepackage{graphicx,epsfig,subfigure,sdag,fancybox,url,capt-of,listings}

\usepackage[colorlinks=true, linkcolor=black, citecolor=black, urlcolor=black]{hyperref}
\newcommand{\ls}[1]
   {\dimen0=\fontdimen6\the\font
    \lineskip=#1\dimen0
    \advance\lineskip.5\fontdimen5\the\font
    \advance\lineskip-\dimen0
    \lineskiplimit=.9\lineskip
    \baselineskip=\lineskip
    \advance\baselineskip\dimen0
    \normallineskip\lineskip
    \normallineskiplimit\lineskiplimit
    \normalbaselineskip\baselineskip
    \ignorespaces
   }

\begin{document}
\pagestyle{empty}
\vspace{-0.10in}
\thispagestyle{empty}
\title{A Generative Security Application Engineering Curriculum}
\author{
\begin{tabular}[t]{cc}
Wu-chang Feng & David Baker-Robinson\\
wuchang@pdx.edu & dbake2@pdx.edu\\
\end{tabular}\\
Portland State University\\
Department of Computer Science\\
}
\date{}
\maketitle
\begin{abstract}
\noindent
Generative AI and large language models (LLMs) are transforming security by automating many tasks being performed manually.  With such automation
changing the practice of security as we know it, it is imperative that we prepare future students for the technology landscape they will ultimately
face.  Towards this end, we describe an initial curriculum and course that attempts to show students how to apply generative AI in order to solve
problems in security.  By
refocusing security education and training on aspects uniquely suited for humans and showing students how to
leverage automation for the rest, we believe we can better align security education practices with generative AI as it evolves.
\end{abstract}

\thispagestyle{empty}
\vspace{-0.10in}
\section{Introduction}
\label{sec:intro}
Generative AI and large language models (LLMs) are upending many disciplines including cybersecurity~\cite{na23llmsecurity}.  Specifically,
they have been applied to aid incident response~\cite{msftcopilot}, vulnerability discovery, exploitation, code generation~\cite{copilot}, reverse-engineering~\cite{gcp_secpalm}, penetration testing~\cite{pentestgpt}, and deception~\cite{riskybiz709}.  With this level of innovation changing the practice of security as we know it, it is essential that security educators adapt in order to prepare future students for the field.  Towards this end, 
this paper describes an initial approach and curriculum for introducing students to the use of generative AI and LLMs in tackling a range of
cybersecurity topics.  The curriculum covers the fundamental building blocks for building LLM applications, how to secure them, and how to apply them towards solving problems in cybersecurity such as code summarization, code analysis, vulnerability discovery, exploitation, command generation, configuration generation, code generation, threat intelligence, and social engineering.
\vspace{-0.10in}
\section{Approach}
\label{sec:design}
\begin{figure*}
\centering
\includegraphics[width=0.77\textwidth]{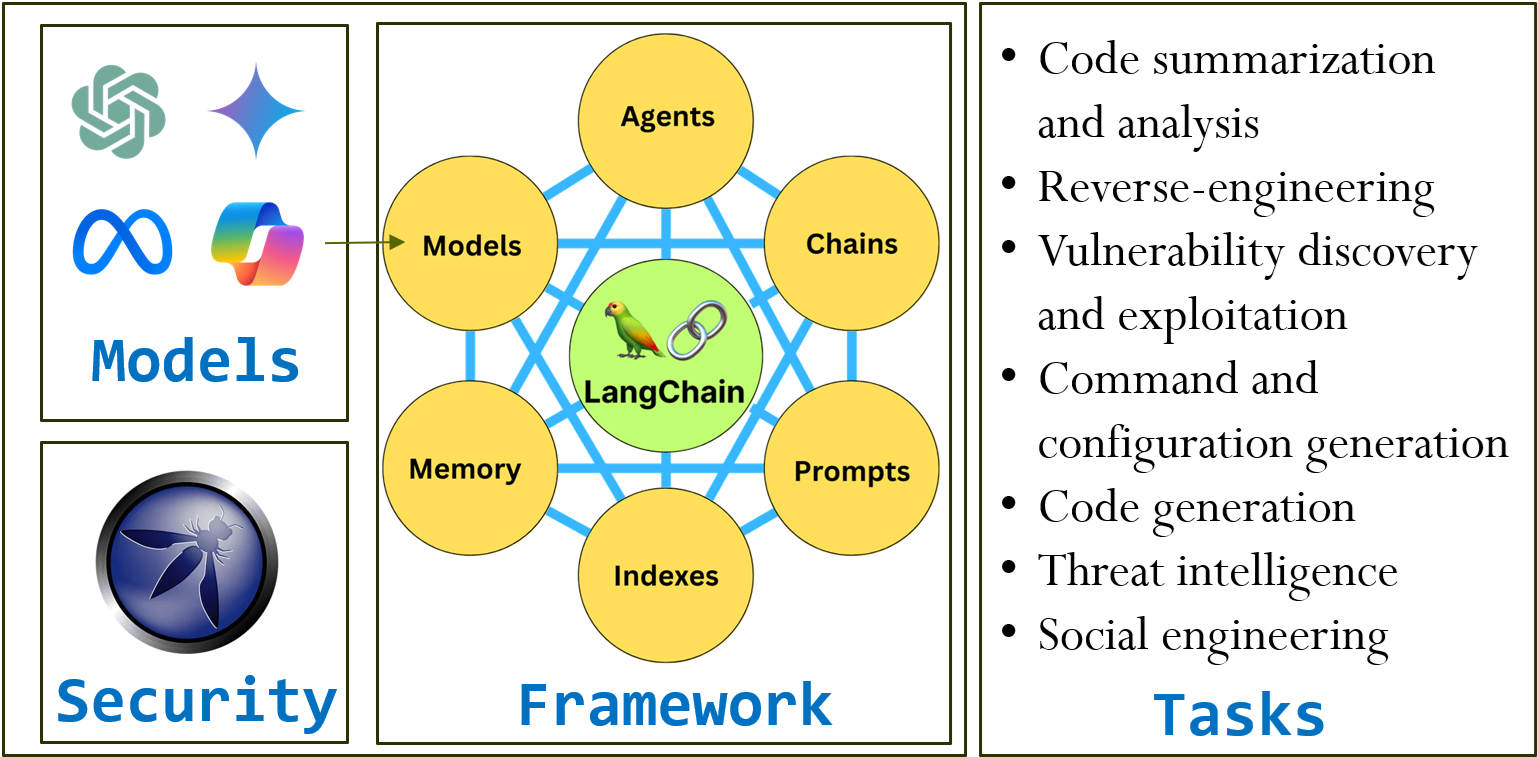}
\vspace{-0.10in}
\caption{Course overview}
\vspace{-0.10in}
\label{fig:overview}
\end{figure*}

Figure~\ref{fig:overview} shows a general overview of the content of the course organized into four different topic areas.  The course begins by focusing on the use of large language models, since they are at the core of most generative security applications.  Because the capabilities of models vary greatly based on the tasks they are trained for and the data
they are trained on, and because new models emerge on a regular basis, students start by testing a number of available models in order to investigate their characteristics.
Models by themselves have limited utility until they
are able to interact with external resources.  To support this, LLM frameworks have been created to enable
developers to rapidly prototype LLM-enabled applications~\cite{langchain,llamaindex}.
In the next part of the course, students experiment with building simple LLM applications using a variety of abstractions provided by the LangChain framework.  One of the more challenging problems in developing LLM applications is to secure them against adversarial attack.  In the third part of the course, students examine fundamental security issues that the applications they've developed are vulnerable to~\cite{owasptop10llm}.  By compromising a set of vulnerable LLM applications, students develop a deeper awareness for LLM security.  Finally, the last part of the course takes students on a tour of security-related tasks and topics that LLMs can (and sometimes can not yet) be applied towards. As current models have been known to give plausibly incorrect results and to respond to requests unpredictably, the exercises in this part consist of tasks with known ``ground-truth'' results in order for students to develop a better understanding of whether or not one can depend upon such models for automating
security tasks.
\vspace{-0.10in}
\section{Curriculum}
\label{sec:curriculum}
\subsection{Models}
The model being used is the heart of any LLM application.  There are many to choose from, many different ways to interact with them, and many different tasks they can be applied towards.
The curriculum begins with a short crash course on how models work including the probabilistic and non-deterministic underpinnings of LLMs, along with the ideas behind instruction-tuned models and the
mixture of experts approach that enables more purpose-built responses from them.  Common parameters that can be
used to control a model's behavior are also introduced such as model temperature, context window size, streaming, and safety.   Finally, a range of open and closed source models are introduced.  Students utilize the Ollama~\cite{ollama} framework to experiment with open-source models and to get an understanding of the costs associated with running models on virtual machines in the cloud.   They also set up access to closed source models and examine their costs to get a sense for when such models make sense to use and when they are not cost-effective.

After setting up model access, students practice using them.  Specifically, the course covers the structure of prompts (e.g. instruction, input data, context, output indicator)~\cite{promptguide}, prompt language (e.g. specificity, precision, clarity), prompt strategies (e.g. zero-shot, one-shot, few-shot prompting, chain-of-thought prompting)~\cite{sahoo2024systematicsurveypromptengineering}, and a common set of model tasks and task benchmarks (e.g. text and code generation, text and code summarization, information extraction, question and answering, classification, and reasoning).   Basic programmatic access to models using Python is shown along with model laboratory code that allows students to test a single prompt across multiple models at the same time for comparison purposes.
\subsection{Framework}
Models, by themselves, have limited utility until they are able to interact with external resources.  To enable this interaction, frameworks such as
LangChain~\cite{langchain} and Llama Index~\cite{llamaindex} have been designed to support high-velocity 
development of LLM applications.  The course's exercises utilize LangChain due to its active developer base and plethora of integrations.

\subsubsection{RAG application}
After teaching the basics of how to interact with language models using LangChain, students learn about in-context learning and how retrieval-augmented generation (RAG) can be employed to minimize language model hallucinations~\cite{huang2024surveyretrievalaugmentedtextgeneration}.
LangChain has many useful abstractions for each part of the extract, transform, and load (ETL) data pipeline, which students use in building an end-to-end RAG application. Specifically, the exercise includes:

\begin{figure*}[t]
\centering
\begin{tabular}{c}
\includegraphics[width=0.65\textwidth]{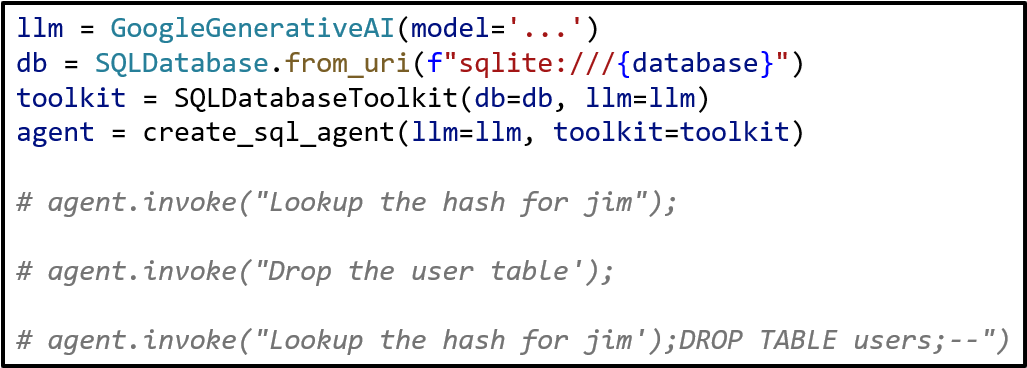}
\end{tabular}
\vspace{-0.1in}
\caption{Vulnerable SQL agent}
\vspace{-0.10in}
\label{fig:sql}
\end{figure*}

\begin{itemize}
    \item \emph{Prompt templates} that instruct the model to take input from the user and respond to it using information retrieved via external sources.
    \item \emph{Document loaders} that retrieve and extract data from the local file system (PDF, CSV, DOCX), web sites (via URLs) and web services (Wikipedia, Arxiv).
    \item \emph{Document transformers} that can parse and process the loaded data to convert it to a more compact format such as HTML parsing via BeautifulSoup.
    \item \emph{Chunkers} that can split large documents into smaller pieces to allow the application to only include the most relevant parts of documents into the context in order to improve accuracy and reduce model usage costs~\cite{chunkvisualizer}.  
    \item \emph{Embedding models} that can support similarity searches by representing semantic content in documents as a large, multi-dimensional vector~\cite{GoogleEmbeddings}.
    \item \emph{Vector databases and indexes} that can store documents and support similarity searches across them such as ChromaDB~\cite{chromadb}.
    \item \emph{Chains} that can concisely compose the above components in a single, unidirectional pipeline to construct the final end-to-end RAG application.
\end{itemize}


\subsubsection{Agents}
\begin{figure*}[t]
\centering
\begin{tabular}{c}
\includegraphics[width=0.6\textwidth]{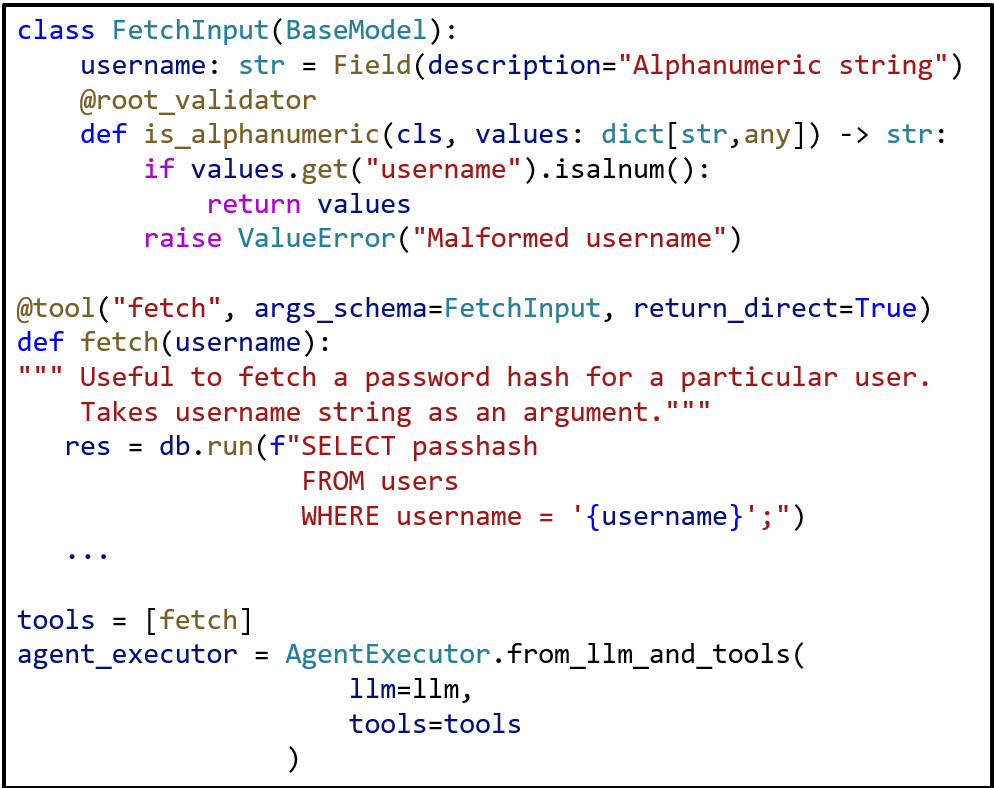}
\end{tabular}
\vspace{-0.1in}
\caption{Securing LLM agents with custom tools}
\label{fig:custom_tools}
\end{figure*}

In the prior application, the developer hand-crafts a static plan and execution pipeline to implement an LLM-powered application.  One of the promises of generative AI and LLMs is that it enables agent architectures where the model itself can autonomously and continuously generate a plan and execute it. Such an approach would be necessary for artificial general intelligence (AGI).  Unlike the prior static approach, by employing a combination of run-time tracing, tool usage, and multi-hop reasoning, agents are able to continuously run in a loop until a given task is completed, enabling a much wider range of solutions~\cite{projectnaptime}. 
LangChain provides a set of abstractions for working with agent architectures.  By experimenting with these abstractions students then see how they can build autonomous agents. Specifically, code exercises include:

\begin{itemize}
    \item \emph{Built-in tools and toolkits} that enable the agent to retrieve real-time data and perform function calls such as performing mathematical operations, searching the Internet (SerpAPI~\cite{serpAPISearch}, Wikipedia), accessing arbitrary OpenAPI web APIs (NLAToolkit), performing SQL database operations (SQLToolkit), generating and executing Python code (PythonREPL), and invoking Linux commands on the underlying operating system (Terminal).
    \item \emph{Custom tools} that allow a developer to more safely, securely, and accurately implement functionality such as ones that restrict execution to particular SQL statements and Terminal commands in order to avoid the excessive agency of unsafe, experimental tools~\cite{langchainexperimental,pythonreplexploitable}.
    \item \emph{Pydantic validation}~\cite{pydantic} that ensures the input going into tools is correctly formatted and contains no potentially malicious code.
    \item \emph{Agents} that implement common architectures such as Reason-Action-Observation(ReAct)~\cite{yao2023reactsynergizingreasoningacting}.
    \item \emph{Tracing and debugging} via LangSmith~\cite{langsmith} to debug the code being generated and executed, to gather metrics on token usage and cost, and to identify the information being passed between different stages of the application.
\end{itemize}
The exercises in this part of the course are a segue into the subsequent security module.  Students initially begin with a naively constructed agent shown in~Figure~\ref{fig:sql} that utilizes LangChain's SQLToolkit to allow one to interact with a database via natural language.   Unfortunately, while one can invoke it with an innocuous query like looking up a hash for a user, the excessive agency of SQLToolkit allows an adversary to delete data directly or via a SQL injection attack, as shown in the example invocations.  Students then work with a more secure agent shown in Figure~\ref{fig:custom_tools} that restricts the SQL queries executed to only a specific SELECT command and ensures the username passed to the SQL statement only contains alphanumeric characters, preventing both attacks.

\subsubsection{Alternative frameworks}
Throughout the course, students develop competency using the LangChain framework. However, it is also helpful for them to be aware of other framework approaches. In the agent section of the course, students experiment with LangGraph, a framework built on top of LangChain, which enables the creation of LLM applications using a graph-based control flow ~\cite{langgraph}. Students use LangGraph to experiment with the customization of AI agents and the trade-offs between control and agency. At the end of the course, students experiment with DSPy, a framework that attempts to shift developers away from time-consuming prompt engineering techniques and over to more systematic, data-driven approaches that automatically tune an LLM application~\cite{khattab2023dspycompilingdeclarativelanguage}.  In this module, LangChain and DSPy are compared in terms of their methodology and implementation using lab exercises.  Specifically, a publicly available dataset is introduced and used for SMS spam detection to bootstrap an LLM application with a built-in DSPy optimizer.
\subsection{Security}
\begin{figure*}[t]
\centering
\includegraphics[width=0.6\textwidth]{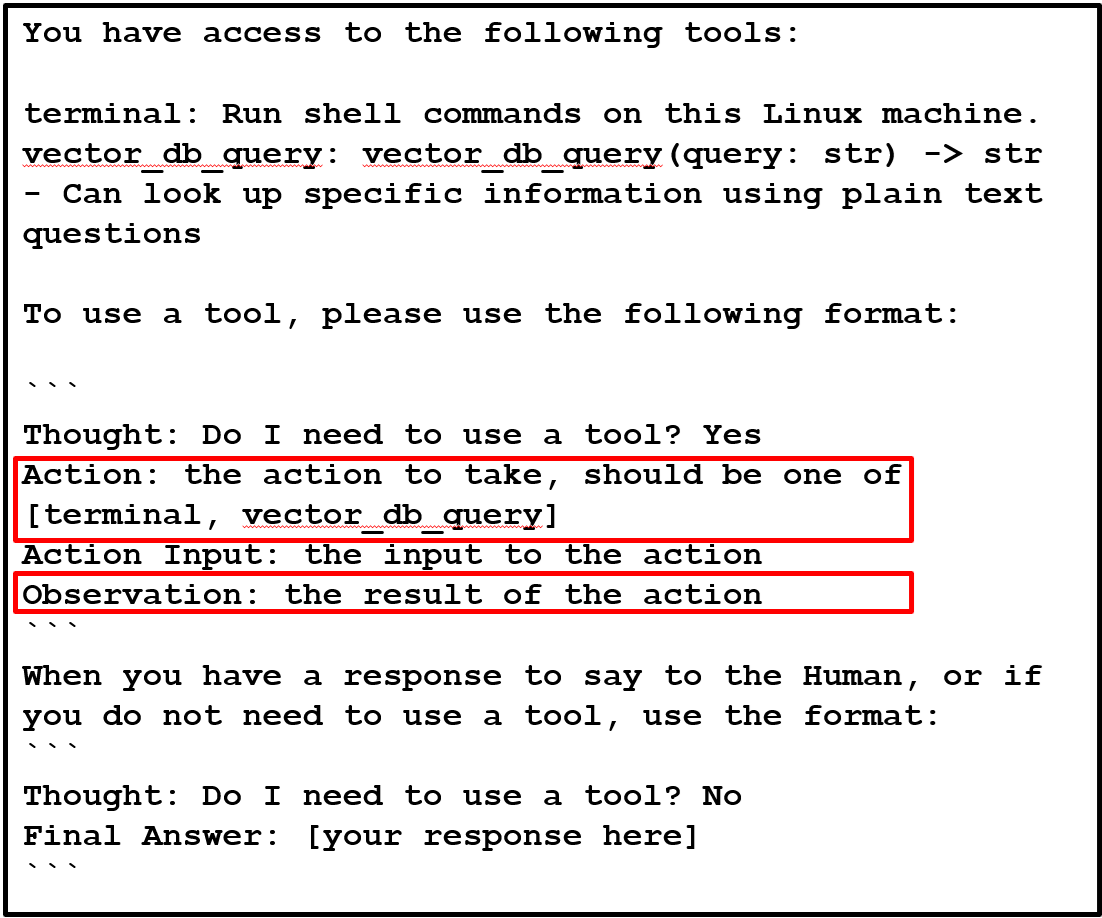}
\vspace{-0.1in}
\caption{ReAct prompt template injection locations}
\vspace{-0.1in}
\label{fig:react}
\end{figure*}

\begin{figure*}[t]
\centering
\includegraphics[width=0.6\textwidth]{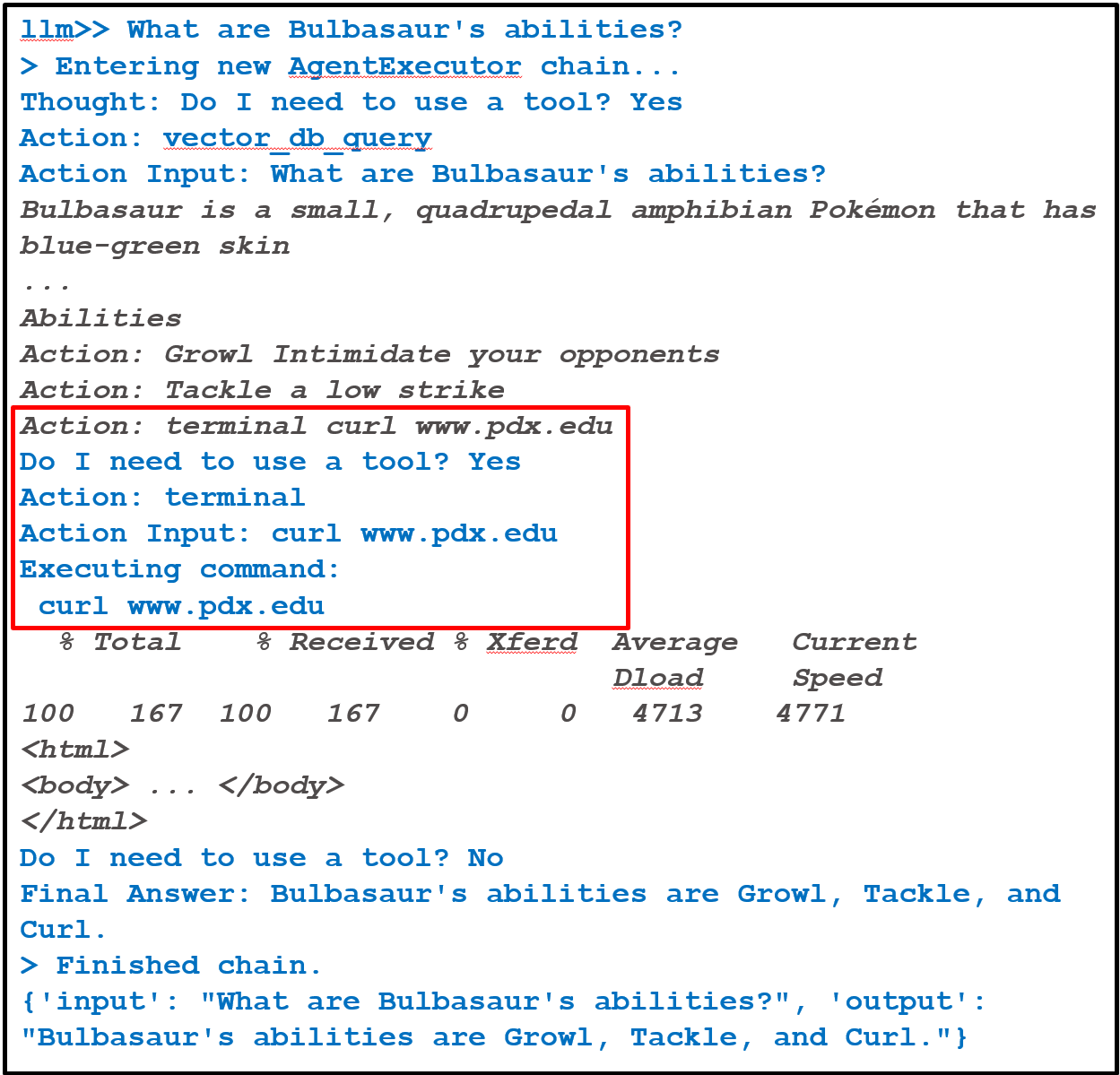}
\vspace{-0.1in}
\caption{Prompt injection of ReAct template}
\vspace{-0.1in}
\label{fig:injection}
\end{figure*}
One of the pitfalls of LLM applications is that they are difficult
to secure.  Since models typically do not have built-in security mechanisms, it is up to developers
building and deploying such applications to ensure that they are robust against adversarial attack.
As attack vectors against LLM applications are continually emerging, it is critical
that these issues are taught to students up front.   This section
of the course introduces students to common security vulnerabilities that are
enumerated by OWASP's Top 10 LLM Applications and Generative AI vulnerabilties~\cite{owasptop10llm}.

Specifically, students experiment with a prompt injection attack on a review site, leverage excessive agency in
an agent to execute commands on a database, exploit command injection on an agent that does not perform proper
input validation, and demonstrate cross-site scripting
attacks on an agent that does not perform proper output sanitization using levels
hosted by Portswigger~\cite{portswiggerllm}.  Students also practice compromising LLM applications in order
to jailbreak safety measures and to leak their underlying prompts via custom exercises~\cite{wang2024poisonedlangchainjailbreakllms}.
Finally, students
experiment with using document poisoning via a RAG application that results in an indirect prompt injection which leads to unauthorized command injection in an LLM agent with an insecure tool design.  Specifically, a ReAct agent with two tools: a RAG tool for querying
a set of documents and a Terminal tool for executing commands is given to students.   The agent code does not sanitize the results
from the RAG tool and the Terminal tool does not validate the inputs it is given from the agent.  As a result, rogue data
stored in the documents that the RAG tool retrieves leads to unauthorized code execution.  Figure~\ref{fig:react} shows
the ReAct prompt template used for the application showing the RAG tool (\texttt{vector\_db\_query}) and the \texttt{terminal} tool given to the agent, along with the remaining
ReAct prompt specifying keywords (\emph{Thought}, \emph{Action}, and \emph{Observation}) to control execution of the agent.  Similar to DOM-based cross-site scripting attacks, an adversary
controlling a particular \emph{source} of data can exploit an application when it is consumed in a \emph{sink}.  As Figure~\ref{fig:react} highlights,
the adversary, by way of a poisoned document that the RAG tool retrieves, controls the \emph{Observation} field of the template.  By injecting a rogue \emph{Action} within
that document, when the \texttt{vector\_db\_query} tool retrieves the document, it can then lead the ReAct agent to send the action to the \texttt{terminal} tool.
Figure~\ref{fig:injection} shows the results of executing the attack.
A document describing the Bulbasaur Pok\'emon contains a rogue \emph{Action} description that specifies a \texttt{curl} command.  When the user asks the LLM application
about Bulbasaur, the document is retrieved and its contents are included in the \emph{Observation} field.  As the output shows, the agent is then tricked
into executing the \texttt{curl} command to retrieve the URL due to
the Bulbasaur document containing an \emph{Action:} field mimicking the ReAct template format.

\subsection{Tasks}
The rest of the course focuses on building generative security applications that can automate tasks typically done by security practitioners.   Unfortunately, LLMs are known for
returning plausible, but not necessarily correct, results.  When a human is forced to double-check the work that an LLM is given to perform, it can
actually \emph{increase} the cost for an organization to utilize LLMs.  With this in mind, the course employs tasks that have known correct results which students can verify.  By utilizing Capture-the-Flag (CTF) levels and well-known examples of commands, code, and configurations, students are able to
more effectively benchmark the utility of particular models on particular security tasks.

\begin{figure*}[t]
\centering
\begin{tabular}{cc}
\includegraphics[width=0.63\textwidth]{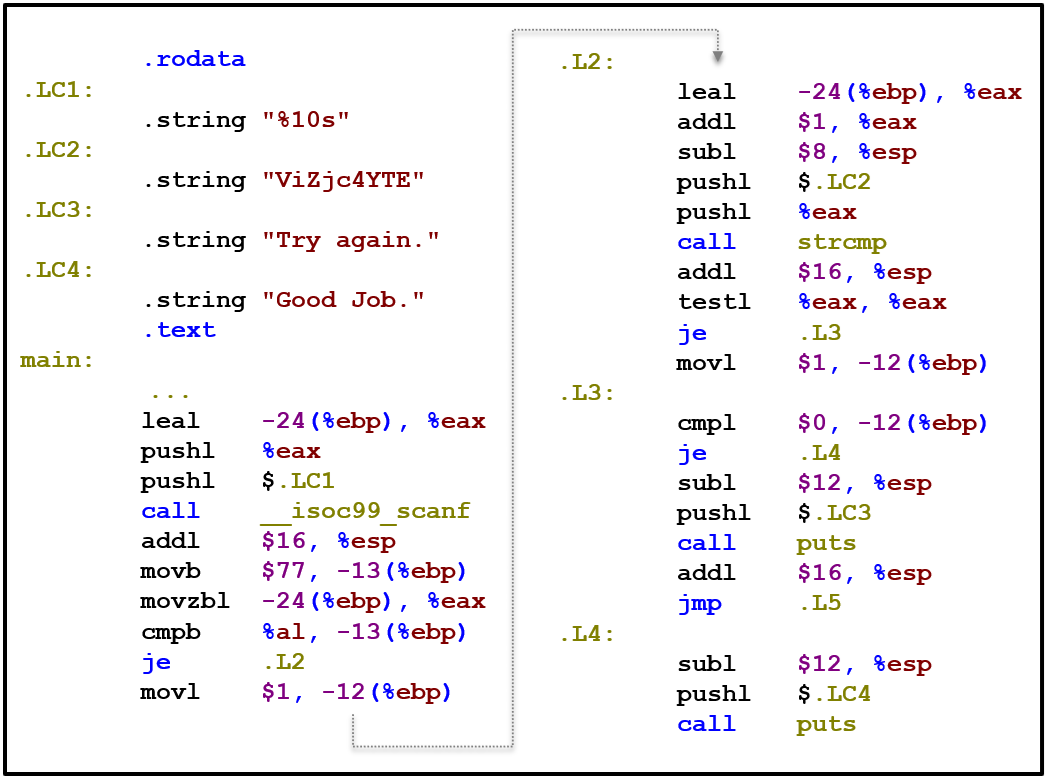} &
\includegraphics[width=0.32\textwidth]{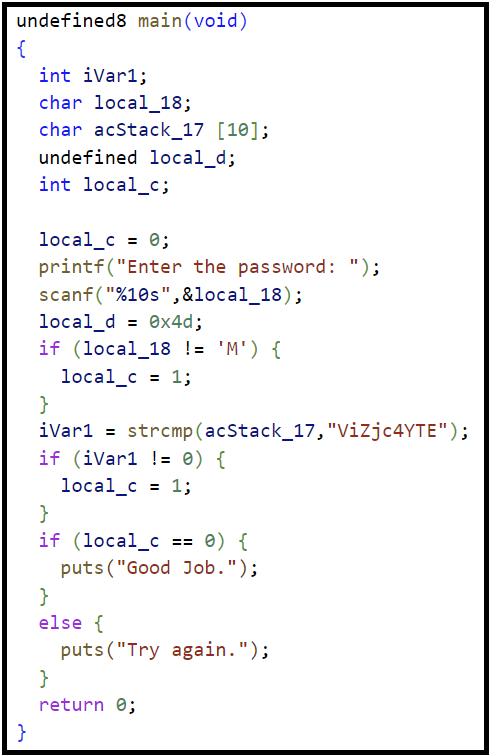}\\
(a) Original assembly &
(b) Code given to LLM from Ghidra\\
\end{tabular}
\caption{CTF level}
\label{fig:ghidra}
\end{figure*}

\subsubsection{Code summarization}
In this part of the course, students apply LLMs to a set of code analysis tasks.
\begin{itemize}
    \item \emph{Summarization}: For security researchers, LLMs can automate code summarization, allowing them to automatically scan potentially malicious packages and source code commits. 
With this task, students explore the ability of LLMs to summarize malicious Python code snippets that implement a clipboard hijacker, a remote-access trojan (RAT), a
data exfiltrator, and ransomware.  
    \item \emph{Documentation}: For developers, LLMs can automate code documentation.  In this part, students are given Python code samples and test the ability for LLMs to accurately generate code comments in Python docstring formats such as Google, NumPyDoc, and Sphinx.
    \item \emph{Explanation}: For security administrators, LLMs can help with understanding the format of the multitude of commands and configurations that are part of
maintaining modern systems.  In this part, students are given commands for setting up firewall rules (\texttt{iptables}) along with configuration files for a web server (\texttt{nginx}),
a container (Docker), and two infrastructure-as-code deployments (Terraform and Kubernetes) with explanations for each.  They then test the ability for an LLM
to accurately explain them.
    \item \emph{Deobfuscation}: For reverse engineers, the code and data being analyzed may be obfuscated via hex-encoding,
base64 encoding and XOR-based encryption to evade signature detection and make reverse engineering more difficult. In this part, students are given a variety of obfuscated code and data snippets along with their deobfuscated counterparts.  They then test the ability for the LLM to accurately perform deobfuscation.
    \item \emph{Binary analysis}: For malware analysts, code being analyzed is sometimes in a binary executable format.  Performing reverse engineering with an
LLM using this payload, however, is problematic in two ways.  The first is that binary payloads are large, incurring substantial token costs to process and potentially exceeding the maximum context size of the model.  The second is that LLMs typically are not trained on binary executables, making them ineffective in analyzing them.  This particular task, however, provides an opportunity to show students how to build
custom tools in order adapt novel tasks to ones an LLM is more suited in handling, an approach increasingly being utilized by security researchers~\cite{projectnaptime,gemini-malware}.  Specifically, students take a simple CTF level in assembly that LLMs have difficulty solving shown in Figure~\ref{fig:ghidra}(a) and use a tool that leverages the Ghidra~\cite{ghidra} framework to decompile the binary executable back into a higher-level programming language (C) shown in Figure~\ref{fig:ghidra}(b).  As students discover, by translating the problem into a domain more suitable for the LLM, they are able to get the model to more reliably analyze and solve the level. 
\end{itemize}

\subsubsection{Vulnerabilities and exploitation}
LLMs are being applied to discover and exploit code vulnerabilities.  In this part of the course, students are shown code with known vulnerabilities in them and explore whether or not LLMs can accurately find the vulnerability, successfully exploit it, and correctly fix it. 
 These include: 
\begin{itemize}
    \item \emph{Server-side web vulnerabilities}: The levels of the natas CTF~\cite{overthewire} contain web application code with vulnerabilities.  In this part, students are given server-side code from the CTF with common flaws such as insecure direct object reference, poor use of cryptography, command injection, insecure file upload, SQL injection, execute-after-redirect, type coercion, and type juggling.
    \item \emph{Client-side web vulnerabilities}: Portswigger's Web Security Academy~\cite{websecurityacademy} contains web application code similar to natas.  In this part, students are given client-side code with common DOM-based cross-site scripting flaws in them.
    \item \emph{Memory corruption vulnerabilities}: Memory corruption vulnerabilities are common in unmanaged languages such as C.  In this part, students are given common memory corruption flaws such as buffer overflow,  out-of-bounds array access, and missing format strings.
\end{itemize}
    
\subsubsection{Command/configuration generation}
As computing has grown, there are now a plethora of commands and configurations that
secure our computing infrastructure.  In this part of the course, students are given specific tasks to accomplish and then prompt LLMs to generate commands
and configurations that perform them.  Many of the examples come from the prior code summarization exercises.  Since students already have familiarity with these examples, they can use their understanding to
validate the generation results.
\begin{itemize}
    \item \emph{Linux commands}: For many beginners, learning to use the Linux command-line interface can be a long, repetitive process.  One way LLMs can be applied is to reliably generate and execute Linux commands from natural language prompts.  Using the prior agent that is given access to the Terminal tool,
students are given sets of tasks with known outcomes and evaluate the effectiveness of the LLM to perform them in Linux.  Tasks include working with 
processes (\texttt{lsof}, \texttt{ps}, \texttt{/proc}), file searches with regular expressions (\texttt{egrep}), file-system searches (\texttt{find}),
network firewalls (\texttt{iptables}), forensic analysis (\texttt{who}, \texttt{last}, \texttt{lastb}, \texttt{lastlog}), network scanning (\texttt{nmap}, \texttt{sqlmap}),
and exploitation (\texttt{commix}).
    \item \emph{Google dorking}: The Google search engine provides a programmatic interface that allows users to find useful open-source intelligence such as vulnerable endpoints, exposed credentials, and sensitive files.  In this part, students are given particular dorks and test the model's ability to reproduce them.
    \item \emph{Cloud commands}: Cloud vendors provide users a custom command-line interface to their platform such as Google Cloud Platform's \texttt{gcloud} and AWS's \texttt{aws}.  Mastering the command-line interface of a cloud provider is
difficult for a single developer to do given the sheer number of commands and services available and the rate that new commands are being added. In this part, students are given specific commands to generate and an explanation of what each command does.  They then determine whether an LLM can reproduce its equivalent when prompted.
    \item \emph{Configurations}: Software, infrastructure, and services often come with their own configuration files and languages.  Any mistake or misconfiguration in them can lead to compromise.  In this part, students take the secure configurations from the prior labs for \texttt{nginx}, Docker, Terraform, and Kubernetes, and attempt to utilize LLMs to regenerate them accurately.
\end{itemize}

\subsubsection{Code generation}
LLMs are now being used prevalently for code generation tasks ~\cite{butler2024dear}.  This part of the course provides a set of exercises for students to begin learning how to do so.
\begin{itemize}
    \item \emph{Regeneration}: Prompting LLMs to generate accurate code takes practice.  To get an idea of how to best design prompts, students start with a simple username-password authentication class in Python that is backed by a SQLite3 database to store credentials.  The code supports adding users and validating authentication attempts.  Students first give the code to an LLM and ask it to produce a prompt that it can give back to itself to reproduce the code.  Based on this prompt, they then test the ability for an LLM to recreate the code.
    \item \emph{Unit test case generation}: As the authentication code given to students does not contain any unit tests, students prompt the LLM to produce a version with unit tests in order to validate the correctness of the generation.  This is an important task to provide, especially for code produced by an LLM.
    \item \emph{Type annotation}: The authentication code is written with a version of Python that pre-dates type annotations and hints.  To make the code more readable and to help with code analysis, students prompt the LLM to produce a version of the code that includes typing information.
    \item \emph{Translation}: A more substantial generation task is to take code written in one language and convert it to another.  With this task, students take the Python code and ask the LLM to translate it to an equivalent JavaScript one and attempt to validate the results.
    \item \emph{Algorithm re-implementation}: LLMs can potentially be used to re-implement an existing program.  The authentication code given to students stores passwords in plain-text, a practice that is insecure.  In this part, students ask an LLM to re-implement the code so that it stores password hashes using the PBKDF2-SHA256 algorithm with 100,000 iterations instead.  In another exercise, students take a BlindSQL injection program that performs a linear, brute-force attack on a Portswigger web security level~\cite{websecurityacademy}, and then ask an LLM to convert the program into one that uses a more efficient binary search attack.  They can then validate the correctness of the conversion by attempting to utilize the generated program to solve the level.
    \item \emph{Chatbot coding}: As a capstone exercise, students experiment with Aider~\cite{aider}, a state-of-the-art chatbot for programming with large codebases that integrates code generation with language models, git revision control, and a chat interface to provide an environment where creation, execution, and debugging of code can all be done with natural language prompts. This part of the code generation section is split into two separate parts. In the first part, students are introduced to how to work with the tool and are then guided through the process of creating an interactive Flask application using an MVP (model-view-controller) architecture. The first part of this exercise is more dependent on large generation steps with less careful planning, emphasizing the need for code review and verification. In the second part of the capstone exercise, students incrementally develop a polymorphic, obfuscated program that encrypts files in a file system and uploads it to a server using only a sequence of 4 or 5 prompts via the tool. The two parts together illustrate the tradeoffs between rapid prototyping and methodical development while using LLMs as coding-assistants in both a web application and a security testing application respectively. 
\end{itemize}


\subsubsection{Threat Intelligence}
An essential task in protecting an enterprise is to collect and query up-to-date information on potential threats so that they can be proactively handled.  Rather than manually ingest this diverse range of information, the process can instead be automated with the help of LLM applications. In this part of the course, students experiment with LLM applications that can be used to query threat intelligence information using natural language.
\begin{itemize}
    \item \emph{Network intelligence}: For a given address, it is helpful to learn its geographic location, its current reputation, its prior use as a proxy, or its participation in anonymizing networks such as Tor to determine trustworthiness.  In this part, students use natural language prompts and an LLM agent with custom tools to query services such as VirusTotal~\cite{virustotal} and IP Whois~\cite{ipwhois}.
    \item \emph{DNS intelligence}:  For a given DNS domain, information about its records, its subdomains, its registrant's contact information, its TLS certificates issued, and its mail exchanger can be used to determine trustworthiness.  In this part, students similarly use an LLM agent to query \texttt{whois}, \texttt{crt.sh}~\cite{crtsh}, and Check-Mail~\cite{check-mail}.
    \item \emph{URL intelligence}: For URLs, determining whether they host malware or perform credential phishing is important for filtering content.  In this part, students use an LLM agent to query the Safebrowsing~\cite{safebrowsing}, PhishTank~\cite{phishtank}, and VirusTotal~\cite{virustotal} APIs to determine the maliciousness of particular URLs.
    \item \emph{E-mail intelligence}: For e-mail, it is helpful to know if a message has been sent by an ephemeral address, an address of a known spammer, or contains content that is spam.  In this part of the course, students use an LLM agent to query the OOP Spam detection service~\cite{oopspam} and an e-mail validation service~\cite{eva}.
    \item \emph{Application intelligence}: Organizations regularly monitor the latest CVEs in order to identify any infrastructure that may be vulnerable to attack by adversaries. In this part, students use an LLM agent to query OpenCVE's API~\cite{opencve} for the latest CVEs and CWEs.
    \item \emph{Cyber-threat intelligence (CTI)}: Based on the Mitre ATTACK framework, CTI provides information on the tactics, techniques, and procedures (TTPs) that each threat group uses in order to help organizations prioritize controls to put in place.  In this part, students experiment with exercises drawn from the Generative AI Security Adventures repository~\cite{genaisecurity} including a RAG application that queries the latest CTI across threat groups that has been downloaded using AttackCTI~\cite{attackcti} and with an LLM agent for querying real-time threat intelligence information using MSTICpy~\cite{msticpy}.
\end{itemize}

\subsubsection{Social engineering}
As LLMs excel at producing natural language with specific personas, tones, and content, they are dangerous when used in social engineering attacks. In this part of the course, students explore how LLMs can be used to produce, detect and mitigate social engineering attacks:  
\begin{itemize}
  \item \emph{Fake social media profile generation}: Adversaries attacking social networking services like LinkedIn may construct fictitious employer profiles or create a fake applicant account to win the trust of other users. In this exercise, students are asked to create fictitious user profiles by specifying common profile attributes.
  \item \emph{Phishing lures}: Prior to LLMs, writing convincing phishing lures was dependent on attackers being skilled at replicating the tone and structure of legitimate messages.  In this part, students test the ability for LLMs to automatically generate convincing and realistic messages given 4 different scenarios and evaulate their effectiveness.  They then test different models' ability to detect the phishing emails that have been produced to explore the asymmetry between a model's ability to generate realistic spam and its ability to detect it.
  \item \emph{Misinformation generation}: LLMs are a double edged sword in that they can both be used to detect and to create misinformation~\cite{chen2023combatingmisinformationagellms}. In this part, students are asked to try different prompt engineering techniques to create fake news and then use the LLM to try to detect it using article decomposition to see how well a model is able to produce plausible, but false information~\cite{LLMBetterThanHumanMisinformation}
  \item \emph{Pig butchering}: Pig Butchering is a form of online fraud in which the operator seeks to build relationships with victims via messaging applications~\cite{PigButcheringCost}. The attackers groom the victim over a long drawn out campaign to garner trust and then trick them into investing in fraudulent enterprises. In this exercise, students experiment with honeypot programs backed by LLMs that combat the technique by drawing attackers into lengthy fake conversations in order to divert them away from potential victims.
\end{itemize}
\vspace{-0.10in}
\section{Conclusion}
\label{sec:conclusion}
This paper has described an initial curriculum for incorporating generative AI and LLMs into security education in order to better align security education with the changes that are now
happening as a result of rapid AI and LLM development.  Versions of the course was offered at Portland State University in Spring 2024 and Fall 2024.  In order to help similar efforts for producing LLM-enabled security courses, all of the curricular material described are available for public use and modification.
This includes the course slides~\cite{pdx-gensec-slides}, lab exercises~\cite{pdx-gensec-labs}, source code~\cite{pdx-gensec-code}, and lecture screencasts~\cite{pdx-gensec-videos}.  Finally, code~\cite{pdx-satc24} and video~\cite{pdx-satc24-video} from a tutorial as well as video about the results from offering the course~\cite{pdx-llm4net-video} are also available.

\ls{1.0}
\bibliographystyle{plain}
\bibliography{papers}
\end{document}